\documentclass[pdflatex,sn-mathphys-num]{sn-jnl}

\usepackage{graphicx}%
\usepackage{multirow}%
\usepackage{amsmath}
\usepackage{amsmath,amssymb,amsfonts}%
\usepackage{amsthm}%
\usepackage{mathrsfs}%
\usepackage[title]{appendix}%
\usepackage{xcolor}%
\usepackage{textcomp}%
\usepackage{manyfoot}%
\usepackage{booktabs}%
\usepackage{algorithm}%
\usepackage{algorithmicx}%
\usepackage{algpseudocode}%
\usepackage{listings}%
\usepackage{siunitx}

\theoremstyle{thmstyleone}%
\theoremstyle{thmstyletwo}%

\theoremstyle{thmstylethree}%

\begin{document}

\title[Article Title]{Free-form diamond refractive optics enable efficient high-energy X-ray nano-imaging}


\author[1,2]{\fnm{Aknur} \sur{Karabay}}\email{aknur.karabay@epfl.ch}

\author[3]{\fnm{Xianbo} \sur{Shi}}\email{xshi@anl.gov}

\author[4]{\fnm{Frank} \sur{Seiboth}}\email{frank.seiboth@desy.de}

\author[5,6]{\fnm{Carlos S.} \sur{Baraldi Dias}}\email{carlos.dias@diamond.ac.uk}

\author[4]{\fnm{Azat} \sur{Khadiev}}\email{azat.khadiev@desy.de}

\author*[4]{\fnm{Nazanin} \sur{Samadi}}\email{nazanin.samadi@desy.de}


\author*[1,2]{\fnm{Manuel} \sur{Guizar-Sicairos}}\email{manuel.guizar-sicairos@psi.ch}

\affil*[1]{
\orgname{Paul Scherrer Institute (PSI)}, 
\orgaddress{\street{Forschungsstrasse 111}, \city{Villigen PSI}, \postcode{5232}, \country{Switzerland}}}

\affil[2]{
\orgname{École Polytechnique Fédérale de Lausanne (EPFL)}, 
\orgaddress{\street{Rte Cantonale}, \city{Lausanne}, \postcode{1015}, \country{Switzerland}}}

\affil[3]{
\orgname{Argonne National Laboratory}, 
\orgaddress{\street{9700 S Cass Ave}, \city{Argonne}, \state{IL}, \postcode{60439}, \country{USA}}}

\affil[4]{
\orgname{Deutsches Elektronen-Synchrotron (DESY)}, 
\orgaddress{\street{Notkestraße 85}, \city{Hamburg}, \postcode{22607}, \country{Germany}}}

\affil[5]{
\orgname{Karlsruhe Institute of Technology (KIT)}, 
\orgaddress{\street{Hermann-von-Helmholtz-Platz 1}, \city{Eggenstein-Leopoldshafen}, \postcode{22607}, \country{Germany}}}

\affil[6]{
\orgname{Current address: Diamond Light Source}, 
\orgaddress{\street{Harwell Science and Innovation Campus}, \city{Didcot}, \postcode{OX11 0DE}, \country{United Kingdom}}}


\abstract{Full-field transmission X-ray microscopy (TXM) enables nondestructive three-dimensional imaging of thick and strongly absorbing materials with high spatial resolution. Such capabilities are essential for understanding structure-function relationships in hierarchical materials, with broad applications in biology, energy conversion, and energy storage. At high photon energies, however, TXM performance is limited by the reduced efficiency of diffractive optics and by the challenge of matching the numerical aperture (NA) of the illumination to that of the objective optics. Here we harness free-form diamond refractive optics to overcome a key illumination-efficiency bottleneck in high-energy TXM, demonstrating full-field nano-imaging at 20 keV with a half-period resolution of 72 nm. The optical system combines a custom-designed diamond refractive beam shaper that produces a uniform near-flat-top illumination at the sample with a 94\% efficiency, a moving diffuser placed near the sample to increase the effective illumination NA and improve image quality and resolution, and high-precision aberration-corrected diamond compound refractive lenses as the objective optics. 
These results establish free-form diamond optics as a powerful route to efficient high-energy TXM, expanding full-field nano-imaging to complex, evolving materials systems and enabling in situ, operando, and tomographic studies under experimentally realistic conditions. Furthermore, it opens new avenues for innovation of joint X-ray optical-digital design for a new generation of high-energy X-ray nano-imaging. }


\maketitle

\section{Introduction}
\label{sec_main}

Full-field transmission X-ray microscopy (TXM) enables nondestructive nanoscale three-dimensional imaging of extended samples across both the soft- and hard-X-ray regimes. At hard-X-ray photon energies, exceeding several keV, TXM provides high-throughput imaging of thick samples  with deep penetration and enables studies in dense or strongly absorbing materials while maintaining high spatial resolution~\cite{sakdinawat2010nanoscale_review}. This has enabled TXM as a particularly attractive tool for the study of technologically and scientifically relevant specimens across a wide range of scientific disciplines~\cite{ Andrews2010-mineral-density-txm, Stampanoni_2010_TXM, Wang2015-dinosaur-teeth-TXM, ZHAO2018batteryTXM, Vincent2021_TXM_sub10nm, Flenner2023, Nikitin2025}. Compared with scanning approaches such as ptychography or scanning transmission X-ray microscopy, full-field TXM offers substantially higher throughput and a simpler data-analysis workflow, as an entire image is recorded in a single shot. An advantage that is especially important for \textit{in situ}, \textit{operando}, and tomographic experiments, where rapid acquisition is required~\cite{Nelson2012, Wang2014-batteries-operando, Wang2014-in-situ-lithium-anodes,Ge2018}. 

While TXM has thrived in quality and resolution at energies below 12 keV, approaching or exceeding 10 nm resolution \cite{Vincent2021_TXM_sub10nm}, the upgrade of synchrotron X-ray facilities to low-emmitance fourth-generation sources significantly increases the available brilliance that can be used for imaging at even higher energies. This new technology opens the door for nanoscale tomography with energies at or beyond 20 keV, with important applications in imaging with nanoscale resolution. Such high photon energies would enable nanoscale tomography of millimeter-range volumes of tissue, imaging deeply buried metallic layers or defects in fully packaged integrated circuits, and access much more complex and extreme \textit{operando} or \textit{in situ} environments, for example battery materials under realistic operating conditions.


A central component of TXM systems is the beam-shaper optics, which shape and redistribute the incident X-ray beam to provide homogeneous illumination over the field of view while matching the illumination numerical aperture to that of the objective~\cite{sakdinawat2010nanoscale_review, Stampanoni_2010_TXM}. 
A variety of beam-shaper concepts have been developed for X-ray microscopy, including Fresnel zone plate (FZPs) condenser~\cite{Anderson2000-circular-zoneplate-condenser}, capillary condensers~\cite{Yin2006-capillary-condenser, Zeng2008-glass-capillary-condenser}, refractive lenses~\cite{Reznikova2007-refractive-condenser}, multilayer mirrors~\cite{Stollberg2006_multilayer_mirrors}, and more complex diffractive beam shapers with linear gratings divided into individual subfields~\cite{Stampanoni_2010_TXM}. Among these, the diffractive beam shaper has become the dominant solution in high-resolution hard-X-ray TXM because of its flexibility in beam shaping, compatibility with zone-plate objectives and flexibility to jointly design Zernike phase plates~\cite{zernike1935phase,sakdinawat2010nanoscale_review}. However, diffractive beam-shaper condensers suffer from the same efficiency limitations as diffractive objectives at high photon energies, particularly when high numerical apertures are required. For high-resolution Fresnel zone plates, the first-order diffraction efficiency at high X-ray energies is typically only on the order of 10\%, with the exact value depending on the choice of zone material and thickness, substrate attenuation, and zone duty cycle. 
Their inherent chromaticity also restricts the usable bandwidth and requires refocusing of the beam shaper when the X-ray energy is changed. 
Alternative reflective capillary condensers exploit grazing-incidence reflection to provide near-unity focusing efficiency and achromatic operation, eliminating the need for refocusing when the photon energy is changed~\cite{Zeng2008-glass-capillary-condenser}. However, their performance becomes increasingly constrained at high photon energies as the critical angle decreases, making large-NA, uniform, and NA-matched illumination difficult to achieve without highly precise capillary geometry and surface quality. 
Moreover, both capillary and diffractive beam shapers require a central stop to block the unreflected or undiffracted light, which further reduces the transmitted flux unless special optics are used to pre-shape the beam~\cite{Samadi2023, Samadi2025}. All these limitations have made high-resolution TXM beyond 15–20 keV challenging, with state-of-the-art half-pitch resolutions of 70 nm at 20 keV and 100 nm at 37.7 keV demonstrated using FZP condensers and objectives \cite{Takeuchi2021_70nm_20kev}.

Current laser ablation technology enables 3D shaping of diamond with sufficient precision for X-ray optics. Recently, a stack of 60 compound refractive lenses was demonstrated to focus an X-ray beam to 52 nm at 14 keV and 68 nm at 20 keV, with efficiencies of 28\% and 46\%, respectively~\cite{Wang25} and similar optics have been used as objective for dark-field X-ray microscopy~\cite{staeck_2026_diamond_dark_field}. The same fabrication approach has also been used to fabricate axicon optics for beam pre-shaping, achieving an efficiency of 83\% at 11 keV~\cite{Samadi2025}. Using diamond refractive optics offers higher efficiency at high X-ray photon energies and greater resistance to radiation-induced damage. 

Here, we leverage this capability to create free-form diamond optics and demonstrate full-field TXM at 20~keV using all refractive, high-efficiency beam shaper and objective. We quantitatively evaluate its photon efficiency, illumination uniformity, and impact on image quality using a lithographically patterned gold test structure. By combining the refractive beam shaper with an upstream movable decoherer and a downstream movable diffuser, we achieve stable, homogeneous flat-top illumination over a 50~µm field of view, enabling high-contrast imaging with a spatial resolution of 72~nm, close to the expected theoretical performance of the refractive objective. These results pave the way for future free-form diamond refractive optics as an efficient approach for nano-imaging in next-generation X-ray sources.

\section{Results}\label{results}

\subsection{The diamond refractive TXM setup}
\label{sec_imaging_setup}

Experiments were performed at the HIKA endstation of the P23 beamline at PETRA III, DESY, Germany. A monochromatic 20 keV X-ray beam was selected using a double crystal Si(111) monochromator. The refractive diamond beam shaper was designed using the subfield-overlap concept commonly used in diffractive beam-shaper condensers. Rather than focusing the incident beam to a single spot, the optic divides the beam into multiple square sub-apertures, each redirecting a portion of the beam to overlap at the sample plane and thereby produce a flat, homogeneous illumination profile.

\begin{figure}[]
\centering
\includegraphics[width=0.9\textwidth]{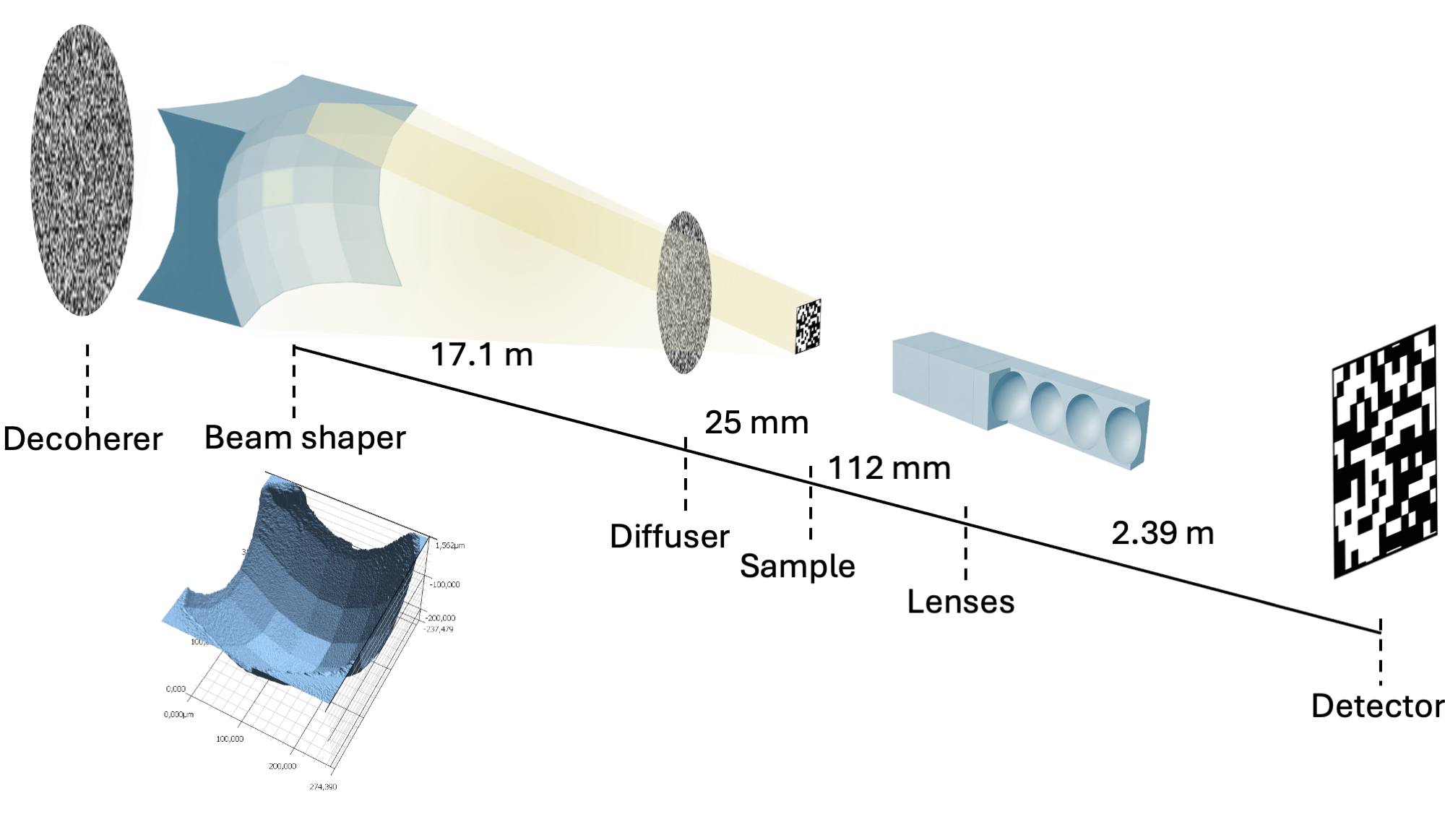}
\caption{Schematic of the high-energy TXM setup using refractive optics. A monochromatic beam is conditioned by a rotating decoherer and incident on the diamond refractive beam shaper. The beam shaper redirects individual beamlets to overlap within the 50~\si{\micro\meter} FOV at the sample plane. A moving diffuser placed immediately upstream of the sample increases the effective illumination NA. A set of diamond refractive lenses is used as the objective, and the magnified image is recorded by the detector. The inset shows the surface profile of one side of the beam shaper measured using a confocal laser microscope (Keyence VK-X1100). }\label{fig_TXM_setup}
\end{figure}

The diamond beam shaper was designed as a two-sided free-form refractive optic fabricated on a 500~\si{\micro\meter}-thick diamond substrate using femtosecond pulsed laser ablation. It consists of a $5\times5$ array of square facets, each with a side length of 50~\si{\micro\meter}, as illustrated in Fig.~\ref{fig_TXM_setup}. Each facet acts as a refractive wedge that redirects an off-axis portion of the incident beam toward the center of the sample plane. As a result, the nearly collimated incident beam is divided into an ensemble of beamlets that overlap at the sample plane and collectively illuminate the TXM field of view (FOV) with a near-flat-top profile. Further details of the beam-shaper design and fabrication are provided in the Methods section.

This free-form refractive design provides considerable flexibility, including the ability to match the beamlet angles to downstream Zernike phase plates~\cite{zernike1935phase,sakdinawat2010nanoscale_review}, which is particularly important for phase-contrast imaging of weakly absorbing materials at high X-ray energies. Unlike diffractive beam-shaper condensers, the refractive beam shaper requires neither a central stop nor an order-sorting aperture, thereby improving the utilization of the incoming beam and simplifying the experimental setup.

Upstream of the beam shaper, the X-ray beam was conditioned by a rotating random decoherer~\cite{Flenner2023}. The decoherer reduces the transverse coherence of the incident beam and suppresses interference between beamlets redirected by the beam shaper, thereby improving the uniformity of the illumination at the sample.

The beam shaper was designed for a source distance of $z_x = 88$~m and a nominal beam-shaper-to-sample distance of $z_s = 21.2$~m. Numerical simulations showed that, for this design, a beam-shaper-to-sample distance of approximately $z_{sc} = 17.1$ m provided a more uniform illumination profile, as shown in the Supplementary Information. This optimized distance was therefore used in the final experimental configuration. The efficiency of the beam shaper was measured to be 94\%, demonstrating that nearly all incident photons within the 250~\si{\micro\meter}~$\times$~250~\si{\micro\meter} acceptance of the beam shaper were redirected onto the sample FOV.

In the present geometry, the diamond beam shaper alone provides a relatively small illumination with $\text{NA}_{illum} = 5.85\times 10^{-6}$, compared with that of the objective $\text{NA}_{obj} = 5.96\times10^{-4}$. For an imaging system, optimal resolution is achieved when the illumination NA is comparable to or larger than the objective NA. When $\text{NA}_{illum} < \text{NA}_{obj}$, the Rayleigh resolution, $\delta x$, can be approximated as \cite{mack2008fundamental_lithography, murphy2012fundamentals_light_microscopy} 
\begin{equation}
    \delta x = 1.22 \frac{\lambda}{(\text{NA}_{illum} + \text{NA}_{obj})},
    \label{eq_resolution}
\end{equation}
where $\lambda$ is the X-ray photon wavelength. Equation (\ref{eq_resolution}) highlights the importance of matching the illumination NA to the objective NA for optimizing spatial resolution, which has driven the challenging design of X-ray beam shapers~\cite{Jochum1995}.  In principle, $\text{NA}_{illum}$ could be increased by stacking multiple beam shapers, analogous to compound refractive lenses, but this would reduce efficiency by adding additional optical elements. Here, we instead used a moving diffuser placed close upstream of the sample. The diffuser introduces angular diversity into the illumination and increases the effective $\text{NA}_{illum}$, providing a better match to the objective $\text{NA}_{obj}$. This improves spatial resolution while further suppressing residual artifacts caused by beam-shaper fabrication defects. 

The wavefield transmitted by the sample propagates downstream to the diamond refractive objective lens assembly. The objective consists of 70 concave diamond lenses arranged in multiple lens cubes (LCs), configured as stacks of 10, 20, and 40 lenses. The lenses were fabricated by femtosecond laser ablation in 3 mm$\times$3 mm diamond substrates with an average thickness of 295~\si{\micro\meter} \cite{Wang25}. 
Wavefront-correcting phase plates, fabricated with a similar technique, were inserted after each 10-lens set. In this configuration, the total length of the assembly was nominally 23 mm, including the phase plates and small air gaps between the mounted LCs. 

Each parabolic lens has a central radius of curvature of 25~\si{\micro\meter} and an aperture diameter of 120~\si{\micro\meter}. A 120~\si{\micro\meter} pinhole was placed upstream of the lenses to define the acceptance aperture and suppress stray light. The full lens assembly had a nominal efficiency of 43.5\% and a focal length of $f=100.7$ mm at an X-ray photon energy of 20 keV. The sample-to-lens and lens-to-detector distances were approximately $z_o = 112$~mm and $z_i = 2.39$~m, respectively. Images were acquired using a pco.edge sCMOS camera with an Optique Peter microscope with 10$\times$ optical magnification, resulting in an effective detector pixel size of 0.65~\si{\micro\meter}. The effective TXM imaging pixel size was measured to be 29.3 nm, corresponding to a TXM magnification of $M = 21.52$.

\subsection{Spatial resolution and imaging quality}\label{analysis}

A high-contrast test sample was used for quantitative image-quality evaluation. The sample consists of a lithographically fabricated gold structure with a characteristic feature period of 10 µm and a thickness of 4~\si{\micro\meter}, providing sharp, well-defined edges for evaluating illumination uniformity, contrast-to-noise ratio, and spatial resolution. 

\begin{figure}[]
\centering
\includegraphics[width=8.5 cm]{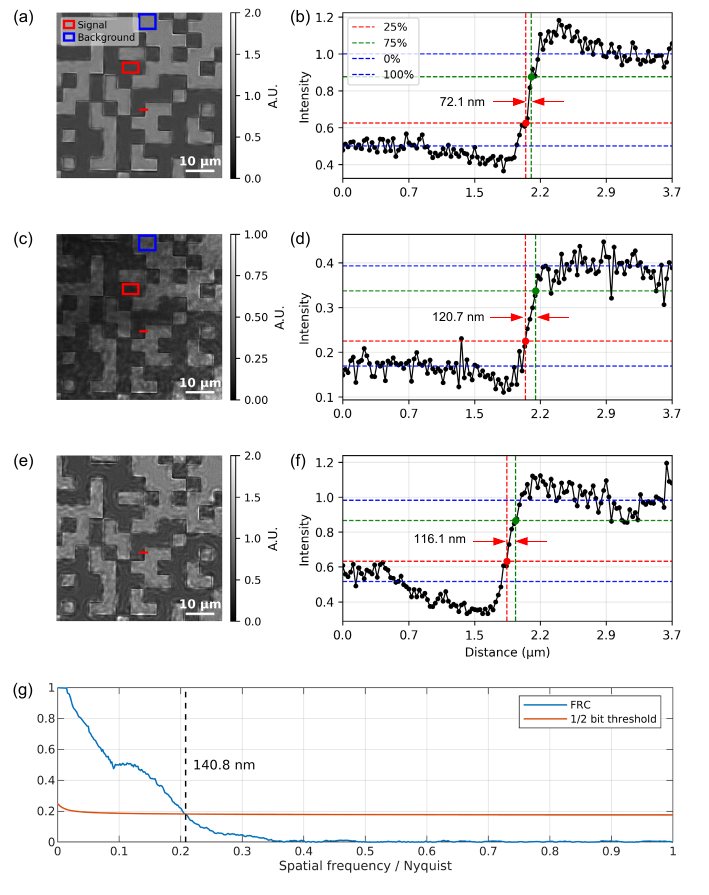}
\caption{Imaging quality and resolution at 20 keV. TXM images of a gold test structure acquired using (a) the complete setup shown in Fig.~\ref{fig_TXM_setup}, and after removing (c) the beam shaper or (e) the moving diffuser near the sample. The corresponding line profiles in (b), (d), and (f) show the spatial resolution estimated using the 25\%–75\% edge-response criterion, with the best half-period resolution reaching 72.1~nm in the complete configuration. Each image is the average of 10 frames with an individual exposure time of 0.1 seconds. Red lines in (a), (c), and (e) indicate the positions of the extracted line profiles. Boxes regions in (a) and (c) indicate areas used for CNR calculation. (g) Fourier ring correlation (FRC) analysis curve showing a half-period resolution of 141 nm.}
\label{fig_images_line_profiles}
\end{figure}

Figure \ref{fig_images_line_profiles}(a) shows an image of the sample resulting from using the complete TXM configuration shown in Fig.~\ref{fig_TXM_setup}. The image was obtained by averaging 10 background- and flat-field-corrected frames, each with an exposure time of 0.1 s. The resulting image exhibits sharp sample features and homogeneous contrast across the gold structures. The line profile shown in Fig.~\ref{fig_images_line_profiles}(b) gives a half-period resolution of 72.1 nm based on the 25\%–75\% edge-response criterion~\cite{Attwood1999_soft_Xrays}. 

For comparison, Fig.~\ref{fig_images_line_profiles}(c) shows an image acquired after removing the beam shaper while keeping both the decoherer and moving diffuser in the beam. As expected, the overall intensity is reduced, and the illumination exhibits stronger spatial inhomogeneity and residual interference patterns, leading to lower contrast and spatially varying background levels. The corresponding line profile in Fig.~\ref{fig_images_line_profiles}(d) gives a degraded resolution of 120.7~nm.

The effect of the moving diffuser was also evaluated by removing the moving diffuser near the sample while keeping the decoherer and beam shaper in place, as shown in Fig.~\ref{fig_images_line_profiles}(e). In this configuration, the beam shaper still provides higher brightness than the case without the beam shaper, but the illumination is less uniform than in the complete configuration. The corresponding line profile in Fig.~\ref{fig_images_line_profiles}(f) gives a resolution of 116.1~nm.

We further evaluated the spatial resolution using Fourier ring correlation (FRC) analysis between two images acquired after translating the sample by 7.37~\si{\micro\meter} transverse to the beam-propagation direction. The FRC analysis, as shown in Fig.~\ref{fig_images_line_profiles}(g), yielded a half-pitch resolution of 141 nm. This value is worse than that obtained from the line-profile analysis, likely because the sample contains relatively large uniform regions with a low density of fine spatial features, which can negatively bias  the FRC estimate, and due to the spatial variations in resolution and image quality which are further discussed below.

Together, these comparisons show that the combination of the decoherer, beam shaper, and moving diffuser provides the most homogeneous square illumination, minimizes intensity modulation across the field of view, and improves both image quality and spatial resolution.

We further assessed image quality using the contrast-to-noise ratio (CNR) \cite{Adler_2004_CNR,Thrane_2017_CNR}. The CNR,  calculated from the boxed regions in Figs.~\ref{fig_images_line_profiles}(a) and \ref{fig_images_line_profiles}(c), is defined as the ratio between the feature contrast and the underlying image noise. For uncorrelated noise, the CNR is expected to increase as $\sqrt{N}$, where $N$ is the number of averaged frames. 

\begin{figure}[]
\centering
\includegraphics[width=8.5 cm]{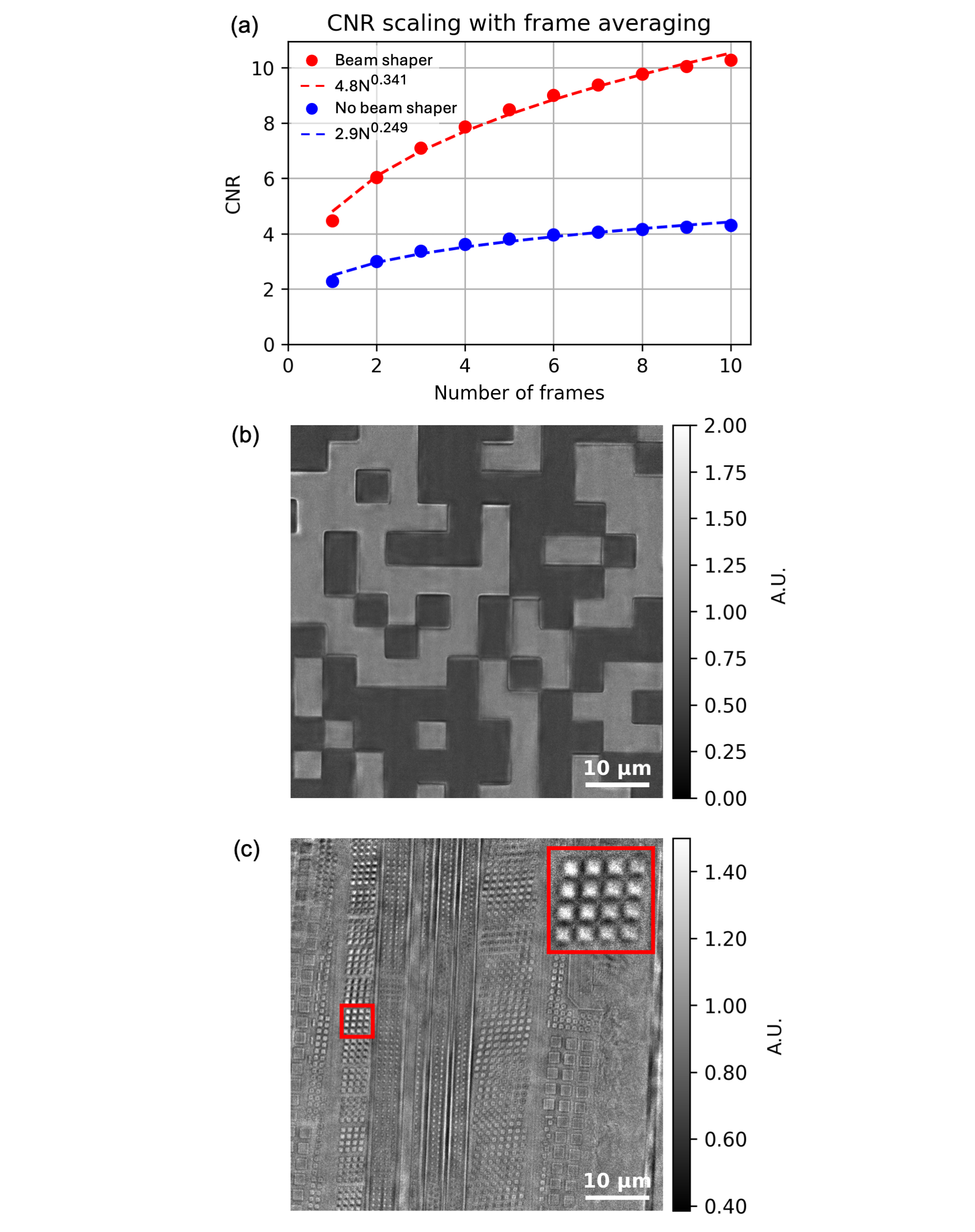}
\caption{(a) Contrast-to-noise ratio (CNR) as a function of the number of averaged frames for images acquired with and without the beam shaper. (b) Reconstructed image obtained by combining two TXM images acquired with a relative sample shift of 7.37~\si{\micro\meter} using Laplacian-pyramid image fusion. (c) TXM image of a TSMC integrated-circuit sample with a 5 \si{\micro\meter}-wide zoom inset.}\label{fig_CNR_fused}
\end{figure}

Figure~\ref{fig_CNR_fused}(a) shows the CNR as a function of the number of frames, each acquired with an exposure time of 0.1~s, for images collected with and without the beam shaper. The use of the beam shaper yields a consistent improvement in the CNR by a factor of 1.96 -- 2.38. Although the fitted power-law exponent does not reach the ideal value of 0.5, presumably due to imaging artifacts, the configuration with the beam shaper shows  a substantially closer exponent value of 0.34. Notably, the CNR achieved with the beam shaper after only two frames already exceeds that obtained without the beam shaper after ten frames, directly demonstrating the gain in imaging throughput and the potential for reduced exposure times and radiation dose.

Notably, spatial variations in resolution and image quality remain visible even in the complete configuration shown in Fig.~\ref{fig_images_line_profiles}(a). These variations are likely caused by accumulated residual surface figure errors across the large number of diamond lenses in the objective assembly. In a single thin imaging lens, such errors mainly act as pupil-plane aberrations and produce nearly field-independent blurring. In a multi-element CRL objective, however, the refracting surfaces are distributed along the optical axis, so figure errors away from the effective pupil plane can introduce field-dependent wavefront distortions, localized blur, and spatially varying image quality. This interpretation is supported by images acquired with a small sample displacement, where the blurred regions remain fixed in the FOV rather than moving with the sample.

Improvements in fabrication are being developed to reduce residual surface figure errors. For this demonstration, we also explored a computational strategy inspired by focus stacking and atmospheric lucky imaging~\cite{law2006lucky,Joshi2010} to create a composite image with improved spatial uniformity. After subpixel image registration~\cite{Guizar2008_subpixel_registration}, the two independent images acquired after a transverse sample shift were combined using Laplacian pyramid fusion~\cite{wang2011}. Because the lens-induced blurred regions remain fixed in the FOV while the sample features shift, the two displaced acquisitions provide complementary sharp regions after registration. The fused image, shown in Fig.~\ref{fig_CNR_fused}(b), effectively suppresses localized lens-induced blur by combining the best-resolved regions from the two acquisitions without introducing visible artifacts. Further details are provided in the Methods section.

Finally, we applied the high-energy TXM configuration to a technologically relevant microelectronic sample to demonstrate its utility for complex, strongly structured materials. Figure~\ref{fig_CNR_fused}(d) shows a TXM image of a TSMC integrated circuit fabricated using a 16~nm technology node. The sample was thinned and polished to a total thickness of 50~\si{\micro\meter} to improve X-ray transmission while preserving a representative multilayer device structure. The 5~\si{\micro\meter} inset reveals fine details of the copper filler patterning, highlighting the ability of the diamond-optics-based TXM system to image buried nanoscale features in complex device architectures at 20~keV.

\subsection{Discussion and prospects for the future}

Here, we have demonstrated a full-field TXM system based entirely on diamond refractive optics, achieving a half-period resolution of 72~nm at 20~keV. The system uses a free-form diamond refractive beam shaper for efficient illumination control and diamond compound refractive lenses as the objective. We further demonstrated the use of a movable diffuser placed upstream of the sample to increase the effective illumination $\text{NA}_{illum}$ and improve matching to the objective NA. Importantly, the refractive beam shaper captures the entire incoming X-ray beam without requiring obscuring elements, such as the central stops needed for other illumination strategies. This all-refractive diamond-optics approach provides an efficient route to full-field nano-imaging above 15~keV, a regime where conventional diffractive-based microscopes are strongly constrained by optical efficiency. By combining high photon energy, high efficiency, and nanoscale resolution, the approach expands TXM toward complex \textit{in situ} environments, buried structures, and strongly absorbing materials at hard-X-ray energies.

\begin{figure}[tp]
\centering
\includegraphics[width=8.3 cm]{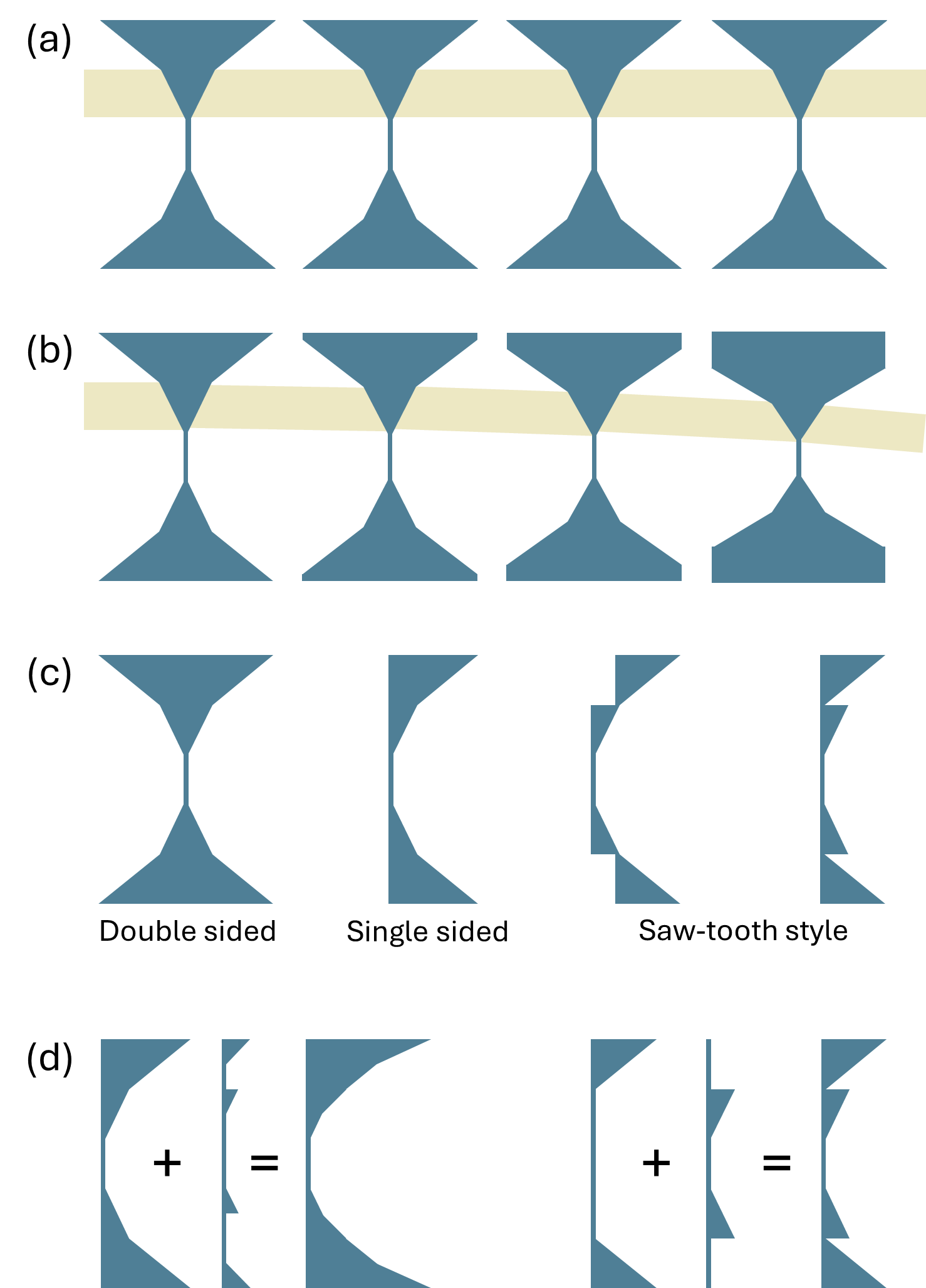}
\caption{\textbf{Compound refractive beam shaper.} (a) Refractive beam shapers can be concatenated in series in order to combine their refractive power and reduce the distance to the sample. (b) For beam shapers with high NA, the design of downstream beam shapers can be adapted to account for the refraction of the beam by the first elements. (c) The experiment described here uses a single double-sided beam shaper, however, single-sided designs are also possible. If steeper angles are feasible for manufacturing, a saw-tooth design can be used to reduce absorption and improve the beam-shaper efficiency. (d) On the left. A modular flexible design platform allows, for example, to take a beam shaper and, by adding a suitable insert reduce the beamlet diameters and increase the NA. On the right, two components can be combined to achieve an equivalent design that would have concave angles otherwise difficult, or even impossible, to manufacture.
}\label{fig_condenser_propects}
\end{figure}

The beam shaper and objective optics were fabricated by femtosecond laser ablation, which offers substantial design flexibility. This approach, for example, has enabled the integration of corrective phase plates into the compound refractive lens assembly used here~\cite{Wang25}, as well as the fabrication of the free-form beam shaper. In the future, this flexibility could support matched designs of beam shapers and Zernike phase-contrast masks. Such design flexibility offers new opportunities in X-ray imaging by incorporating advances in wavefront coding and point spread function engineering to achieve extended depth of field or improved axial sectioning~\cite{Dowski_1995_cubicmask_eDOF,Prasanna_2009_rotating_psf}. Its compatibility with lens-cube mounting also enables straightforward stacking of multiple beam shapers to increase $\text{NA}_{illum}$, forming compound refractive beam shapers, as shown in Fig.~\ref{fig_condenser_propects}(a). Alternatively, a cascade of geometrically scaled beam-shaper elements could be designed, with each downstream element reduced in transverse size to match the shrinking beam footprint after the preceding element, akin to adiabatically focusing lenses \cite{Schroer_2005_adiabatically}, as shown in Fig.~\ref{fig_condenser_propects}(b). 

Improvements in fabrication and surface polishing are underway, which are expected to improve image quality and enable further optics design optimization. For example, higher efficiency can be achieved by replacing  monolithic profiles with sawtooth-style designs that reduce the total diamond thickness traversed by the X-rays, as shown in Fig.~\ref{fig_condenser_propects}(c). 

High-resolution laser ablation imposes a constraint on the maximum manufacturable sidewall angle, because the focused laser beam requires sufficient angular clearance to access the surface. As a result, very steep walls are difficult to fabricate reliably, and the surface slope must remain below a process-dependent limit. Modular compound designs can relax this constraint by distributing the optical function across multiple substrates. As illustrated in the left inset of Fig.~\ref{fig_condenser_propects}(d), an additional refractive element can simultaneously reduce the beamlet size and increase the effective $\text{NA}_{\text{illum}}$.The right inset shows a sawtooth design with concave corners that may be difficult or impossible to fabricate monolithically, but can be simplified by splitting the design across multiple substrates.

Beyond enabling new scientific applications, this flexible design space points toward a new generation of joint X-ray optical–digital design, where illumination, aspherical objective optics, and computational reconstruction are co-optimized across the complete imaging system.

\renewcommand{\theequation}{A\arabic{equation}}
\setcounter{equation}{0} 

\renewcommand{\thefigure}{A\arabic{figure}}
\setcounter{figure}{0} 

\section{Methods}

\subsection{Design and fabrication of the condenser beam shaper}

\begin{figure}[tb]
\centering
\includegraphics[width=8.3 cm]{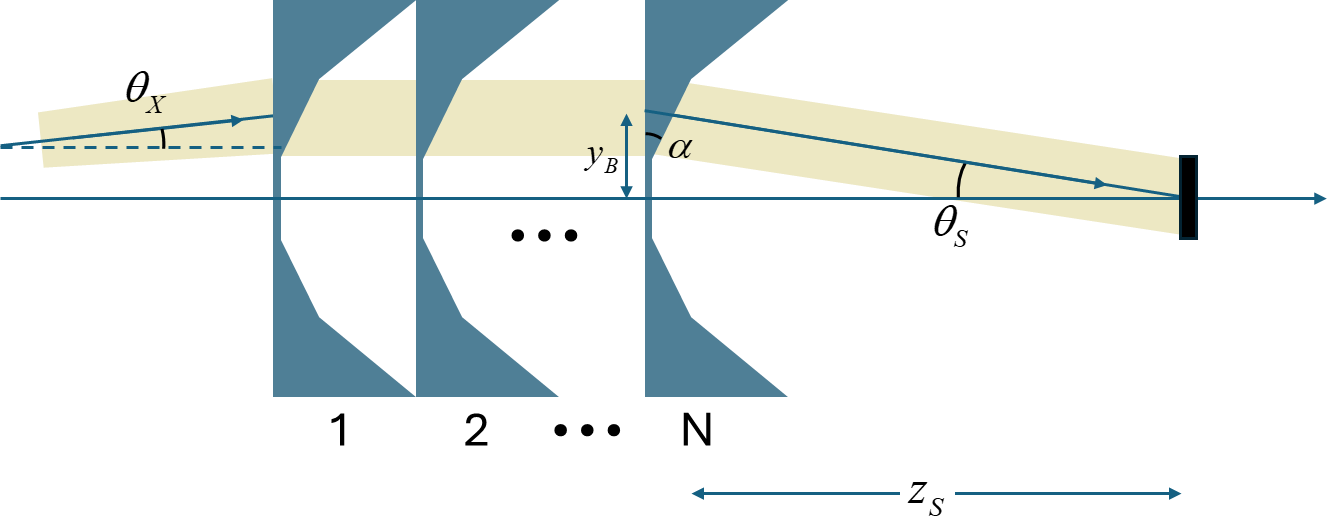}
\caption{Refractive beam-shaper design schematic.
}\label{fig_condenser_design}
\end{figure}

Within the thin-object approximation, the elements of the compound beam shaper satisfy 
\begin{equation}
    \sum_{i=1}^N \tan \alpha_i = \frac{1}{\delta} \bigl( \sin\theta_X +\sin\theta_S \bigr),
\end{equation}
where $\alpha_i$ is the wedge angle of the $i$-th beam-shaper element, as shown in Fig.~(\ref{fig_condenser_design}), $\delta$ is the real part of the refractive index of the beam-shaper material, $N$ is the number of individual beam shapers, $\theta_X$ and $\theta_S$ are the angles subtended by the beam-shaper element as seen by the source and the sample, respectively, as given by
\begin{eqnarray}
    \sin\left(\theta_X\right) = \frac{y_B}{\sqrt{y_B^2 + z_{X}^2}}, 
    \sin\left(\theta_S\right) = \frac{y_B}{\sqrt{y_B^2 + z_{S}^2}},  
\end{eqnarray}
where $y_B$ is the distance of the center of the beam-shaper window to the optical axis, $z_X$ is the distance from the X-ray source to the beam shaper,
and $z_{S}$ is the distance from the beam shaper to the sample. If all the elements of the compound beam shaper are equal, this results in
\begin{equation}
    \tan\left(\alpha\right) = \frac{1}{N\delta}\bigl( \sin\theta_X +\sin\theta_S \bigr).
\end{equation}

The beam shaper and lenses were fabricated at the Centre for X-ray and Nano Science (CXNS), DESY, Hamburg, Germany, using femtosecond pulsed laser ablation. Single-crystal CVD diamond plates were used, and processing was performed with an Amplitude Satsuma HP2 ytterbium-doped fiber laser operating at a fundamental wavelength of 1032 nm and a pulse duration of 285 fs, enabling high precision  surface figuring~\cite{Wang25}. The lenses were fabricated with a parabolic shape, with an aperture diameter of 120~\si{\micro\meter} and a center radius of curvature of 25~\si{\micro\meter}. The double-sided beam shaper, $N=2$, with a total aperture of 250\si{\micro\meter}$\times$250\si{\micro\meter}, was designed as an array of 5$\times$5 beam-shaper elements with individual side-length of 50~\si{\micro\meter}, and distances from the source and the sample of $z_{X} = 88$~m and $z_{S} = 21.2$~m, respectively.

\subsection{Beam shaper efficiency}

Horizontal and vertical slits were used upstream of the beam shaper to achieve a defined illumination of 250\si{\micro\meter}$\times$250\si{\micro\meter} that matched the beam-shaper aperture. To measure the efficiency, an sCMOS pco.edge camera with a 20x objective lens and the effective pixel size of 0.325~\si{\micro\meter} was used to measure the beam with and without the beam shaper, at 1.78 m from the latter. The data was dark-field corrected, and an efficiency of 94.25\% was calculated by the ratio of the total intensity with and without the beam shaper, compared to the theoretically expected value of 97.73\%.

\subsection{Beam-shaper gain}

Table~\ref{tab:condenser_gain} summarizes the intensity gain provided by the beam shaper for different illumination configurations in the experiment. The gain was measured by averaging 10 bright-field acquisitions, \textit{i.e.} without the sample, using the TXM setup shown in Fig.~\ref{fig_TXM_setup}. Each image was measured with 1 second exposure time, and the average was dark-field corrected. The process was repeated with and without the beam shaper. The average intensity was subsequently obtained and the gain was calculated as the ratio of the average intensities with and without beam shaper.

The highest gain was observed for the beam shaper only and beam shaper+diffuser cases, reaching about 13.7$\times$ and 13.2$\times$, respectively, close to the value of 17.74 obtained via numerical simulations. While the decoherer is important to reduce structure in the beam at the sample plane it also caused a decrease in the beam-shaper gain, presumably due to having too high diffusing power. For the complete setup in Fig.~\ref{fig_TXM_setup} the gain was 3.71, this can be improved in future implementations by more judicious choice of the decoherer structure.

\begin{table}[h]
\caption{Beam-shaper gain for different illumination configurations.}
\label{tab:condenser_gain}
\begin{tabular*}{0.5\textwidth}{@{\extracolsep\fill}lc}
 Configuration & Gain \\
\midrule
Beam shaper & 13.73 \\
Beam shaper + diffuser & 13.24 \\
Decoherer + beam shaper & 3.71 \\
Decoherer + beam shaper +  diffuser & 3.71 \\
\end{tabular*}
\end{table}

\subsection{Data analysis}

All images were first subjected to flat-field correction prior to further processing. The corrected intensity was obtained by subtracting the dark-field image, \textit{$I_d$}, from the raw image, \textit{$I$}, and normalizing by the difference between the flat-field image, \textit{$I_f$}, and the dark-field image, as given by
\begin{equation}
\label{eq:ff_correction}
I_{corr} = \frac{I - I_{d}}{I_{f} - I_{d}},
\end{equation}
which compensates for spatial non-uniformities in the illumination.

\subsection{Contrast-to-noise ratio}

The image quality was further assessed using the contrast-to-noise ratio (CNR), which characterizes the contrast of a feature relative to fluctuations in the background~\cite{Adler_2004_CNR}. For this calculation, two rectangular regions of interest (ROIs) from the test structure, representing the signal and background respectively, are used, as illustrated in Fig.~\ref{fig_images_line_profiles}.

Let $I(x,y)$ denote the image intensity at pixel $(x,y)$. We define the signal ROI as $\Omega_{s}$ and the background ROI as $\Omega_{b}$. The CNR is defined as \cite{Osteras2016, muhogora2008}
\begin{equation}
CNR
=
\left|
\frac{\mu_{s}-\mu_{b}}
{\sqrt{\frac{1}{2}\left(\sigma_{s}^{2}+\sigma_{b}^{2}\right)}}
\right|,
\label{eq:cnr}
\end{equation}
where $\mu_{s}$ and  $\mu_{b}$ are the mean intensities in $\Omega_{s}$ and $\Omega_{b}$, respectively, and $\sigma_{s,b}$ denote the signal and background standard deviation, given by
\begin{equation}
\sigma_{s,b} =
\sqrt{
\frac{1}{N_{pix,s,b}-1}
\sum_{(x,y)\in\Omega_{s,b}}
\left[I(x,y)-\mu_{s, b}\right]^2
},
\end{equation}
and $N_{pix, s}$ and $N_{pix, b}$ denote the number of pixels in the signal and background ROIs, respectively.

For the CNR scaling analysis, Eq.~\eqref{eq:cnr} was evaluated for images obtained by averaging different numbers of frames, allowing the improvement in image quality with increased statistics to be quantified for cases with and without using the beam shaper.

\subsection{Effective pixel size calibration}

The test sample has a known feature side length of 5~\si{\micro\meter} which  results in distinct diffraction orders upon Fourier transformation of the image. The accurate location in reciprocal space of those diffraction orders can then provide an accurate estimate of the pixel size. The image in Fig.~\ref{fig_images_line_profiles}(a) was first apodized and zero padded by a factor 10$\times$. Upon Fourier transformation, the image zero-padding results in a 10$\times$ upsampled FT, which allows to estimate the center of the diffraction order with high precision. The pixel size was then estimated by the distance in reciprocal space between the origin and the third diffraction order in the vertical direction, resulting in an estimated pixel size of 29.323 nm

\subsection{Resolution estimation}

The spatial resolution of the TXM images was estimated using two complementary approaches: real-space edge analysis based on line profiles and Fourier ring correlation (FRC) analysis in reciprocal space~\cite{VANHEEL2005}. The line-profile analysis is sensitive to local edge sharpness and provides an intuitive real-space metric, whereas the FRC analysis reflects the reproducibility of the spatial frequency content of the image.

For the real-space estimate, line profiles were extracted across sharp edges of the lithographically defined gold test structure, taken perpendicular to edges as indicated in Fig.~\ref{fig_images_line_profiles}. The minimum and maximum intensities, $I_{\min}$ and $I_{\max}$, were estimated. The 25\% and 75\% intensity levels were defined as
\begin{equation}
I_{25} = I_{\min} + 0.25\,(I_{\max} - I_{\min}), \qquad
I_{75} = I_{\min} + 0.75\,(I_{\max} - I_{\min}),
\end{equation}
and the spatial resolution was determined from the distance between the corresponding crossing points at 25\% and 75\% intensities~\cite{Attwood1999_soft_Xrays}. This yields resolutions of 116.1~nm for the decoherer + beam shaper configuration, 120.7~nm for the decoherer + diffuser, and 72.1~nm for the decoherer + beam shaper  + diffuser.

For FRC resolution estimation, two statistically independent images of the same sample region were acquired with a lateral offset of 7.37~\si{\micro\meter} and subsequently aligned to small fraction of a pixel~\cite{Guizar2008_subpixel_registration}. FRC evaluates the normalized cross-correlation between two independent images in Fourier space over rings of constant spatial frequency. The crossing of the FRC curve with the half-bit threshold \cite{VANHEEL2005} gives an estimated half-pitch resolution of 141 nm for the TXM configuration with decoherer + beam shaper + diffuser. Presumably worse than the line-profile estimates due to the heterogeneous image resolution and due to the sparsity of high-resolution features of the test structure, which have reduced spatial-frequency content and cause a negative bias in the FRC curve.

\subsection{Laplacian pyramid image fusion}
\label{subsec:laplacian}

The fusion process was performed using Laplacian pyramid image fusion~\cite{wang2011,wang_2023_mfif_laplacian_github}, a method developed for combining multifocus datasets by finding the sharpest regions in each image. In this case two registered input images, $I_1$ and $I_2$ acquired with a 7.37~\si{\micro\meter} relative sample displacement to shift the location of lens-induced artifacts were decomposed into their respective Laplacian pyramids. Each pyramid level represents a specific range of spatial frequencies. For a decomposition into $N=6$ levels, the fusion was performed independently at each scale using a set of decision rules tailored to the information content of that band. 

For the upper $N-1$ levels containing high-frequency edge information and fine structural textures, a maximum regional energy rule was applied. For every pixel the local energy $E$ was computed within a $3 \times 3$ sliding window $\omega$ such that $E = \sum_{(x,y) \in \omega} |p_{x,y}|^2$, where $p_{x,y}$ is the pixel intensity. The fused coefficient was selected from the source image exhibiting the highest local response, thereby preserving the sharpest features and highest contrast from each dataset.

For the coarsest low-frequency level containing the global intensity distribution, a weighted information rule was utilized. This rule evaluated both the local variance and Shannon entropy within the neighborhood. In regions where one source image significantly outperformed the other in both metrics, its coefficients were selected; otherwise, a local average was performed to ensure a seamless transition and stable background levels.

The final fused image was reconstructed by the iterative expansion and summation of the fused pyramid levels. As shown in Fig.~\ref{fig_CNR_fused}(b), this image fusion effectively reduced the ``blurry patches'' associated with localized  aberrations without introducing ghosting artifacts. This digital enhancement yields a composite image that represents the peak performance of the objective lens across the entire FOV.

\bmhead{Acknowledgments}
 TXM experiments were performed at the P23 beamline and the HIKA end-station, DESY, Germany. The beam shaper was fabricated at CXNS, DESY, Germany. A.Ka. is supported by ESKAS Swiss Government Excellence Scholarship (2024.0479) and by the the Ministry of Trade, Industry \& Energy Technology Innovation Program RS-2024-00419426, South Korea. Part of this work was funded by Helmholtz Imaging, a platform of the Helmholtz Information \& Data Science Incubator, in the context of a visiting scientist position of M.G.-S.

\bmhead{Author Contributions}
X.S., N.S. and M.G-S. conceived the project and together with C.S.B.D. designed the experiment. M.G-S. designed the diamond beam shaper. F.S. fabricated and imaged the beam shaper and provided the diamond objective lenses. A.Ka., C.S.B.D., A.Kh., N.S., and M.G.-S. carried out the experiment. X.S. and M.G.-S. performed numerical simulations. A.Ka, and M.G.-S. analyzed the data and wrote the paper with contributions from all authors. All authors contributed to the interpretation of the results, read, and approved the manuscript.

\bmhead{Declarations}
F.S. has co-filed a pending patent on 1 March 2024 entitled ‘A lens assembly, a lens unit holder for a lens assembly and a method for providing a lens unit for the lens assembly’, published as EP24160923.9. Applicant for this patent is Deutsches Elektronen-Synchrotron DESY, and the inventors are Frank Seiboth and Ralph Dohrmann. All other authors declare no competing interests.

\bmhead{Data availability}
Data will be made available upon publication of the manuscript in a freely accessible repository.

\bmhead{Code availability}
Codes will be made available upon publication of the manuscript in a freely accessible repository.

\bigskip

\bibliography{sn-bibliography}

\end{document}